\documentclass[twocolumn,secnumarabic,amssymb,nobibnotes,aps,pre,superscriptaddress]{revtex4-1}
\pdfoutput=1
\usepackage{graphicx} 
\usepackage{hyperref}
\usepackage[english]{babel}
\usepackage{datetime}
\usepackage{verbatim}
\usepackage{amsmath}
\usepackage{amsfonts}
\usepackage{amssymb}
\usepackage{siunitx} 
\usepackage{nccmath}
\usepackage{dcolumn}
\usepackage{lineno}
\usepackage{xcolor}
\usepackage{color}
\usepackage{bm}
\usepackage{flexisym}
\usepackage{booktabs}
\usepackage{longtable}
\usepackage{epstopdf}
\usepackage{wrapfig}
\usepackage[utf8]{inputenc}
\usepackage{comment}
\usepackage[normalem]{ulem}

\hypersetup{colorlinks,bookmarksopen,bookmarksnumbered,
citecolor=blue,
linkcolor=black,
pdfstartview=green,
urlcolor=blue}

\addto\captionsenglish{}

\graphicspath{{Figures/}}

\definecolor{green}{rgb}{0,0.5,0}

\def\ee{\end{equation}}
\def\be{\begin{equation}}
\def\bmul{\begin{multline}}
\def\emul{\end{multline}}
\def\bea{\begin{eqnarray}}
\def\eea{\end{eqnarray}}

\newcommand{\Eqref}[1]{Eq.~\eqref{#1}}
\newcommand{\fref}[1]{Fig.~\ref{#1}}

\def\ndclin{N_{\rm disc}} % number of disclinations
\def\ndisloc{N_{\rm disl}} % number of dislocations

\def\qPhi{\Phi} % dislocation flux
\def\qeff{q^{\rm eff}}% effective deficit angle

\def\rhoclin{\rho_{\mathrm{disc}}} % disclination density
\def\rhosloc{\rho_{\mathrm{disl}}} % dislocation density

\def\nD{n_d}    % dislocation number parameter
 
\def\Estretch{E_{\mathrm{stretch}}} % stretching energy
\def\Ecore{E_{\mathrm{core}}} % core energy

\def\cT{c_s}

\def\dA{{\rm d}A\,} % area element
\def\asph{Q}

\makeatletter
\let\cat@comma@active\@empty
\makeatother

\begin{document}

\title{Dislocation screening in crystals with spherical topology}

\author{Ireth Garc{\'i}a-Aguilar}
\affiliation{Instituut-Lorentz, Universiteit Leiden, P.O. Box 9506, 2300 RA Leiden, Netherlands}
\author{Piermarco Fonda}
\affiliation{Instituut-Lorentz, Universiteit Leiden, P.O. Box 9506, 2300 RA Leiden, Netherlands}
\affiliation{Theory \& Bio-Systems, Max Planck Institute of Colloids and Interfaces, Am M\"uhlenberg 1, 14476 Potsdam, Germany}
\author{Luca Giomi}
\email{giomi@lorentz.leidenuniv.nl}
\affiliation{Instituut-Lorentz, Universiteit Leiden, P.O. Box 9506, 2300 RA Leiden, Netherlands}

\date{\today}

\begin{abstract}
Whereas disclination defects are energetically prohibitive in two-dimensional flat crystals, their existence is necessary in crystals with spherical topology, such as viral capsids, colloidosomes or fullerenes. Such a geometrical frustration gives rise to large elastic stresses, which render the crystal unstable when its size is significantly larger than the typical lattice spacing. Depending on the compliance of the crystal with respect to stretching and bending deformations, these stresses are alleviated by either a local increase of the intrinsic curvature in proximity of the disclinations or by the proliferation of excess dislocations, often organized in the form of one-dimensional chains known as ``scars''. The associated strain field of the scars is such to counterbalance the one resulting from the isolated disclinations. Here, we develop a continuum theory of dislocation screening in two-dimensional closed crystals with genus one. Upon modeling the flux of scars emanating from a given disclination as an independent scalar field, we demonstrate that the elastic energy of closed two-dimensional crystals with various degrees of asphericity can be expressed as a simple quadratic function of the screened topological charge of the disclinations, both at zero and finite temperature. This allows us to predict the optimal density of the excess dislocations as well as the minimal stretching energy attained by the crystal.
\end{abstract}

\maketitle

\section{\label{sec:introduction}Introduction}

Crystalline monolayers endowed with spatial curvature are ubiquitous in hard and soft matter across a vast range of length scales: from nanoscopic twisted graphene sheets \cite{Cao2018} to rafts of millimeter-sized soap-bubbles \cite{Bowick2008}. Among all possible crystal structures in two dimensions, the triangular lattice is the most efficient arrangement of particles with isotropic interactions and, as such, its mechanical properties have been studied in a plethora of different systems, including the long-standing Thomson problem \cite{Perez-Garrido1997,Bowick2006,Bowick2002}, viral capsids \cite{Caspar1962,Aznar2012}, colloidosomes \cite{Bausch2003,Einert2005,Lipowsky2005,Irvine2010,Guerra2018}, Abrikosov vortices in thin-film superconductors \cite{Dodgson1997}, Pickering emulsions \cite{Vogel2015} and, more recently, in surfactant-stabilized emulsions \cite{Denkov2015,Guttman2016-1}. 
Crystalline monolayers with spherical topology (i.e. with no boundaries nor handles) form an especially interesting class of two-dimensional partially ordered structures, owing to the impossibility of tiling the sphere with regular hexagons. This results in the appearance of 
disclinations, i.e. lattice sites for which the local coordination number $z$ is different than six. The departure from the ideal six-fold coordinated crystal is quantified via the topological charge $q=6-z$.  
By virtue of Euler's formula, the total topological charge of any triangular lattice constrained on a closed surface is fixed and proportional to its Euler characteristic $\chi$, namely: $Q=\sum_{i=1}^{N}(6-z_{i})=6\chi$, where $N=\sum_{z=2}^{\infty}N_{z}$ is the total number of lattice sites and $N_{z}$ the number of sites of coordination number $z$. For spherical topology in particular, $\chi=2$, and such a constraint is generally fulfilled by introducing a certain number of $5-$fold and $7-$fold disclinations, such that $N_{5}-N_{7}=12$, within an arbitrary number of $6-$fold coordinated lattice sites. The presence of disclinations locally breaks the lattice symmetry group and induces a geometric frustration that leads to non-trivial ground state structures \cite{Giomi2007,Irvine2010,Bowick2009,Manoharan2015}. 

Spherical crystals featuring a low density of lattice sites preferentially organize in icosadeltaheadral structures \cite{Caspar1962}, consisting of $N_{5}=12$ isolated disclinations positioned at the vertices of a regular icosahedron \cite{Dodgson1996,BNT2000}. Such a configuration maximizes the distances between the disclinations, thus minimizing the induced stress resulting from the distortion of the crystal. In denser spherical crystals, or in less regular geometries, defect structures are generally more involved as a consequence of the complex interplay between the distribution of topological charges and the underlying curvature \cite{Nelson1983,Nelson2002Defects}. In particular, two stress screening mechanisms have been extensively studied for spherical geometries: out-of-plane deformations, leading to changes in curvature \cite{Kohyama2007}, and in-plane deformations, with the creation of topologically neutral defects \cite{Perez-Garrido1997,Bausch2003}.

The first mechanism is understood through the fundamental connection between Gaussian curvature and the breaking of orientational order \cite{Nelson1983,Nelson2002Defects}. It is well-known that out-of-plane deformations can promote the formation of defect structures in otherwise regular lattices \cite{Jimenez-Lopez2015}. Conversely, local intrinsic curvature around a disclination can compensate for the angular deficit resulting from having $z_{i}\neq 6$ \cite{Seung1988}. For spherical lattices, buckling around the topological defects leads to the global transformation of the crystal into an icosahedral shape \cite{Lidmar2003, Nguyen2005,Siber2006,Kohyama2007}.

The second mechanism is the spontaneous unbinding of dislocations, whose effect is to delocalize the net topological charge of isolated disclinations, leading to an overall stress relief. Dislocations can be regarded as a tightly-bound pair of disclinations that carries no net topological charge. At equilibrium, they are found in the vicinity of the disclinations or other large lattice distortions, arranged in lines of alternating coordination number. These structures are known as \textit{grain boundaries} on planar crystals \cite{Villeneuve2005} and as \textit{scars} on curved surfaces \cite{Perez-Garrido1997, Bausch2003,Irvine2010}. Contrary to dislocations on a plane, scars are found to terminate within the crystal. Experiments \cite{Einert2005,Guerra2018} and simulations \cite{Kohyama2007,Wales2009,Roshal2014} of closed crystals have revealed that, at finite temperature, most dislocations surround disclinations and form extended defect structures, of size much larger than the lattice spacing. Theoretical models of crystals with uniform curvature have shown that the topological charge, sourcing the elastic stress, is effectively lowered in these extended defects \cite{Travesset2003,Azadi2016}.

These two mechanisms can act separately or simultaneously  depending on the ratio between the local radius of curvature $R$ and the lattice spacing $a$. Spherical crystals with a low density of sites (i.e. with $R/a$ of order one) usually exhibit only twelve isolated disclinations and have been observed to buckle into icosahedral shapes, thus benefiting from curvature screening \cite{Lidmar2003}. Conversely, large crystals with $R\gg a$, tend to have  scars. In fact, seminal experimental work \cite{Bausch2003} has shown that scars only appear at some critical crystal size where then, the number of excess dislocations scales linearly with size, consistent with theoretical predictions \cite{BNT2000,Bowick2006}. 

In the present work, we study how curvature and scars are both involved in the screening of disclination-induced stresses, and how these mechanisms are influenced by the crystal size. Previous numerical works \cite{Gompper1997,Kohyama2007} using particle-based simulations of triangular lattices with spontaneous creation and annihilation of defects, have reported a variety of post-buckling scenarios featuring extended disclination-dislocation complexes, whose position and structure is correlated with the underlying Gaussian curvature \cite{Kohyama2007, Funkhouser2013}.  
These studies, however, have been restricted to particles number in the order of colloidal assemblies. For denser crystals, simulations quickly become computationally demanding and are thus unable to grasp the equilibrium configuration of large structures. To overcome this limitation, here we adopt a continuum approach, where the crystal ground state energy is calculated within the formalism of classical elasticity theory. 

Typically, in continuum models, disclinations are treated as discrete point-like sources of stress and the crystal elastic energy depends uniquely upon their spatial arrangement, the underlying Gaussian curvature and the system Young modulus \cite{Bowick2009}. For spherical crystals with relatively few defects, this approach has proved to be remarkably successful \cite{Bowick2006}, but it becomes increasingly challenging with a growing number of excess defects. Moreover, crystal dislocations may also be entropically generated. Here we take a step further and model scars in terms of a smooth vector field, keeping note of the topological constraint on the dislocation charge. The fact that icosahedral shapes are observed in a large range of scales \cite{Dubois2001,Guttman2019,Guttman2016-1} is an indication that the repulsion between disclination cores is always present, regardless of the presence of scars. In the following, we refer to these twelve topologically required $5-$fold disclinations as {\em seed disclinations} and we will assume they are fixed in the icosahedral configuration. With this and a few more simple assumptions, we are able to show that the stretching energy of a curved crystal with spherical topology of size $R$ takes the general quadratic form 
\be
\Estretch = \frac{1}{2}\,Y a^{2} 
\left[
c_0\,\left(\frac{R}{a}\right)^2
+
c_1 \nD\,\frac{R}{a}
+ 
c_2 \nD^2
\right]\nonumber\,,
\label{eq:introenergy}
\ee
where $Y$ is the Young modulus and $\nD$ is a dimensionless parameter proportional to the number of excess dislocations. The coefficients $c_{1}$, $c_{2}$ and $c_{3}$, which we refer to as the \textit{geometric coefficients}, depend on the specific crystal shape and the distribution of the disclinations and scars within the lattice. Interestingly, this expression disentangles the dependence of the energy on the lattice architecture from the scar density and crystal size. We emphasize that \Eqref{eq:introenergy} is valid for any two-dimensional crystal of spherical topology. The calculation of the stretching energy for a specific lattice architecture is reduced to calculating the geometric coefficients, for which we have developed a numerical method. This method has the advantage of being scale-free, which allows a relatively easy implementation of additional energy terms. 

The remainder of this article is organized as follows. In Sec.~\ref{sec:charge_density} we develop a continuum description of the cloud of dislocation scars emanating from the seed disclinations and we provide generic expressions for the geometric coefficients. In Sec.~\ref{sec:screening} we apply this theory to three specific crystals geometries with spherical topology and various degrees of asphericity.
In Sec.~\ref{sec:finite_temperature} we incorporate into the picture entropically generated dislocations and demonstrate that the stretching energy preserves the generic expression given above. Finally, Sec.~\ref{sec:conclusions} is devoted to conclusions.

\section{\label{sec:charge_density}Defect charge density in large crystals}

\subsection{\label{sec:general_theory}General theory}                     

Let us consider a crystalline monolayer with triangular lattice structure and lattice spacing $a$, on a closed surface of size $R$. Since we deal with closed surfaces, we define the size from the enclosed volume $V$ as $R = [V/(4\pi/3)]^{1/3}$, so that for a sphere $R$ corresponds exactly to its radius.
The elastic energy consists of two terms, penalizing stretching and bending deformations respectively: i.e. $E=E_{\rm stretch}+E_{\rm bend}$. The relative magnitude of these deformation modes is quantified via the F\'oppl-von K\'arm\'an number ${\rm FvK}=YR^{2}/k$, with $k$ the bending rigidity. As demonstrated by Lidmar {\em et al}. \cite{Lidmar2003} in the context viral capsids, while the sphere remains the lowest energy configuration for ${\rm FvK} \lesssim 10^{2}$, for larger ${\rm FvK}$ values the crystal buckles and becomes faceted. Since we are interested in large crystals, we restrict our discussion to minimizers of the stretching energy, using the framework of linear continuum elasticity theory. Here we introduce the main equations, but a more complete derivation and additional details can be found e.g. in Refs. \cite{Nelson2002Defects,Seung1988}.
 
For small deformations, the stretching energy is quadratic in the local strain field $u^{ij}$. At equilibrium, the stress tensor $\sigma^{ij}$ must be covariantly divergence-free i.e. $\nabla_{j}\sigma^{ij}=0$, with $\nabla_{j}$ indicating covariant differentiation.
In particular, if the in-plane strain originates from a generic distribution of disclinations and dislocations, the latter condition can be cast into a Poisson equation for the dimensionless scalar field $\sigma=\sigma_{i}^{i}/Y$, where $\sigma_{i}^{i}=g_{ij}\sigma^{ij}$ is the trace of the stress tensor \cite{Bowick2009}, with $g_{ij}$ the metric tensor of the surface. Namely:
\be
\nabla^2\sigma
=
\rhoclin
+
\rhosloc
-
K
\,,
\label{eq:laplace_sigma_rhos}
\ee
where $\nabla^{2}=g^{ij}\nabla_{i}\nabla_{j}$ is the Laplace-Beltrami operator, $\rhoclin=\rhoclin(\bm{r})$ and $\rhosloc=\rhosloc(\bm{r})$ are respectively the disclination and dislocation charge densities and $K=K(\bm{r})$ is the Gaussian curvature at the point $\bm{r}$ on the surface of the crystal. 
Note that from a geometric perspective, \Eqref{eq:laplace_sigma_rhos} is purely intrinsic and insensitive to bending contributions in the total energy. 

The presence of lattice defects results in singularities in the displacement field $\bm{u}$. In particular, dislocations introduce a discontinuity in $\bm{u}$, disrupting the long-range translational order by an amount characterized by the Burger vector $\bm{b}$. Considering a dislocation as a tightly bound pair of disclinations, this vector has a magnitude of the order of the lattice spacing $a$. By contrast, single disclinations disrupt the bond angle field $\theta$ of the six-fold coordinated crystal by an amount equal to $(\pi/3)q$, with $q=6-z$ the topological charge. Explicitly, the two defect charge densities can be written as discrete sums in the corresponding number of defects \cite{Seung1988}
\begin{subequations}\label{eq:rhos}
\begin{align}
  \rhoclin(\bm{r}) &= \frac{\pi}{3}\sum_{\alpha=1}^{\ndclin} q_\alpha\,\delta(\bm{r}-\bm{r}_\alpha)\,, \label{eq:rhos_discl} \\ 
  \rhosloc(\bm{r}) &= \nabla\times \sum_{\beta=1}^{\ndisloc} \bm{b}_{\beta}\,\delta(\bm{r}-\bm{r}_\beta)\,, \label{eq:rhos_disloc}
\end{align}
\end{subequations}
where $\ndclin$ is the number of the single disclinations and $\ndisloc$ is the number of dislocations. The Dirac delta functions in the expressions above are normalized by the determinant of the metric tensor in such a way to preserve their unit norm when integrated over the surface: $\int \dA\delta(\bm{r})=1$. Note that the curl of a two-dimensional vector is a pseudo-scalar. In a closed crystal with spherical topology, the Gauss-Bonnet theorem implies that the integral of $K$  over the whole surface must be equal to $4\pi$, which in \Eqref{eq:laplace_sigma_rhos} is entirely compensated by the presence of disclinations. In other words, the integral of $\rhosloc$ over the whole crystal must be zero.

For an incompressible crystal, i.e. with unit Poisson ratio, the stretching energy depends only on the Young modulus and can be written in general as \cite{BNT2000}
\be
\Estretch = \frac{1}{2}Y \int \dA \sigma^2 + \Ecore
\label{eq:Estretch}
\ee
where $\Ecore$ is the the defect core energy resulting from the regularization of the continuum theory at distances of the order of the lattice spacing $a$. In practice, this term can be safely neglected as $\Estretch \gg \Ecore$ for $R\gg a$ \cite{BNT2000}.

\begin{figure}
\centering  
\includegraphics[width=0.9\columnwidth]{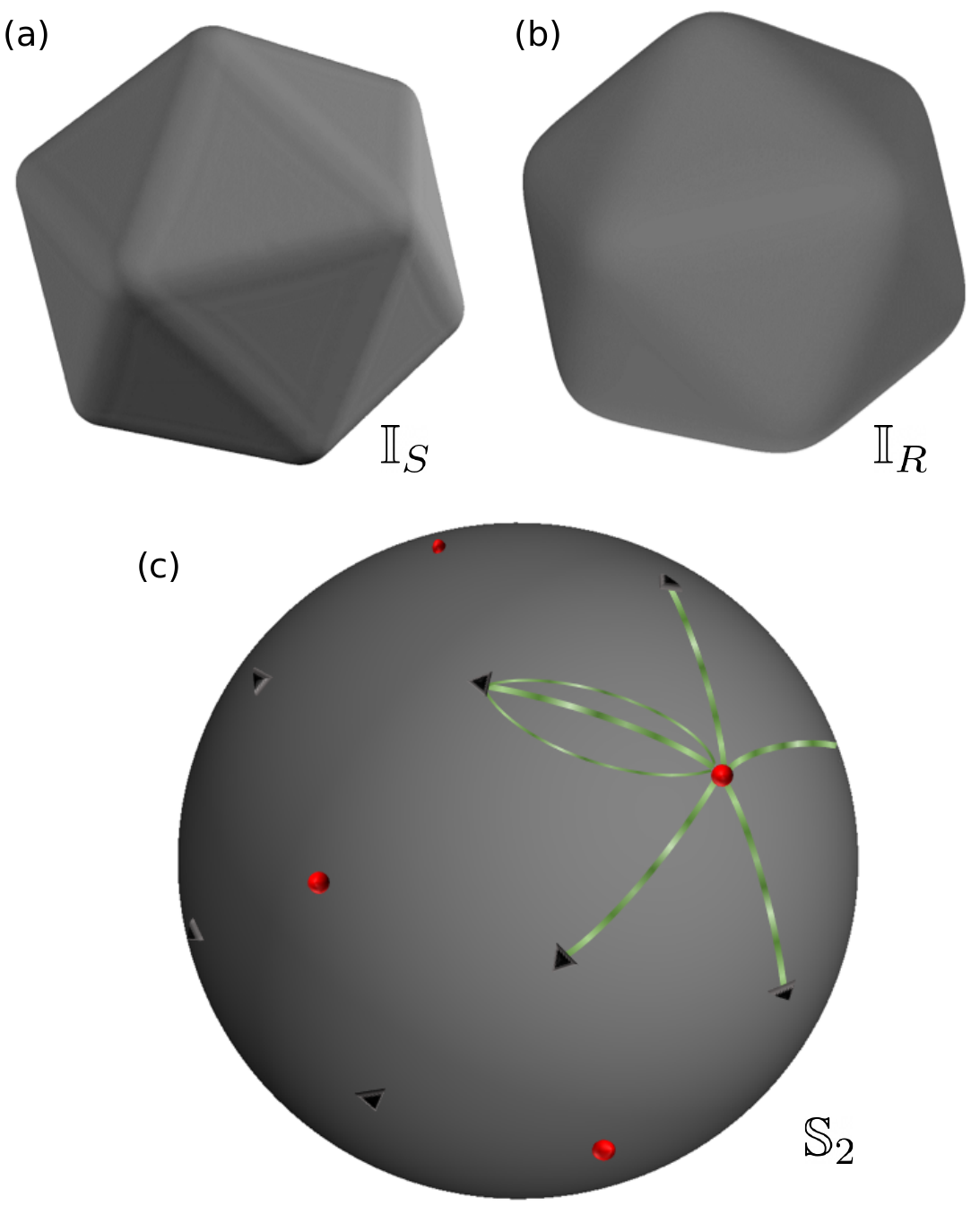}
\caption{
\textbf{Continuum model of screening dislocations on crystals of spherical topology}.  We study the energy of three different crystal shapes  where the local Gaussian curvature around the topological disclinations is varied: the sphere $\mathbb{S}_{2}$, a rounded icosahedron $\mathbb{I}_R$ and a sharper icosahedron $\mathbb{I}_S$. The degree of roundness is characterized by the asphericity $\asph$ defined in \Eqref{eq:asphericity}. On the sphere, we show the schematic of our model for topological defects screening: the scars (green lines) radiate from sources at the seed disclinations (red circles) and terminate at sink positions located at the vertices of a dodecahedron (black triangles).
}
\label{fig:ModelScheme}
\end{figure}

In the following sections, we will construct energy-minimizing defect configurations on various surfaces with spherical topology. In particular, in Sec.~\ref{sec:screening} we focus on three representative crystal shapes: the sphere $\mathbb{S}_{2}$, a round icosahedron $\mathbb{I}_R$ and a sharp icosahedron $\mathbb{I}_S$ with different curvature at the vertices (see \fref{fig:ModelScheme}). We fix the value of the lattice constant $a$ and assume that any two crystals have the same size $R/a$ when they have the same volume 
(see e.g. Refs. \cite{Siber2006,Funkhouser2013,Guttman2016-2}). 
Note that $\mathbb{I}_S$, although sharper than $\mathbb{I}_R$, is not a \textit{perfect} icosahedron, for the principal curvatures would diverge at edges. We characterize the curvature differences between the shapes indirectly through a single value of their asphericity $\asph$, defined in \Eqref{eq:asphericity}, measuring the mean square deviation of the radial distance from a perfect sphere, which clearly has $\asph_{\mathbb{S}_{2}} = 0$. We have $\asph_{\mathbb{I}_{R}} \simeq 0.0012$ for the rounded icosahedron and $\asph_{\mathbb{I}_{S}} \simeq 0.0020$ for $\mathbb{I}_S$ for the sharper one. For reference, with this definition a perfectly sharp icosahedron would have $\asph \simeq 0.0026$. Further details on the characterization of the shapes and how we construct these are found in Appendix~\ref{app:geometries}.

\subsection{\label{sec:effective_defect_charge}Effective defect charge}

As reviewed in the Sec. \ref{sec:introduction}, low-density spherical crystals consists of $N_{5}=12$ isolated $5-$fold disclinations, positioned at the vertices of a regular icosahedron, embedded in an arbitrary $N_{6}=N-12$ number of $6-$fold coordinated lattice sites \cite{Perez-Garrido1997,BNT2000}.
When $R/a \sim 10$, however, scars start to appear \cite{Bausch2003}. Experiments and simulations have shown that scars emanate from single disclinations and are typically oriented towards the center of the triangle formed by three neighboring disclinations \cite{Bausch2003, Guerra2018}. Differently from infinite planar-like crystals, scars terminate within the crystal, thus forming extended defect structures around each disclination which are isolated from one another \cite{Roshal2014,Guerra2018}. For large crystals, it is expected that each dislocation line becomes relatively straight \cite{Perez-Garrido1997} with even several scars branching out as the size approaches the thermodynamic limit \cite{Wales2009}. Based on this evidence, we assume that in the limit of dense crystals, scars will emanate radially out of the twelve seed disclinations and terminate somewhere on the surface away from the icosahedral vertices. In our model we refer to these terminating lattice positions as scar \textit{sinks} (see \fref{fig:ModelScheme}). Since icosahedral crystals result from curvature screening around the disclinations \cite{Kohyama2007,Lidmar2003}, we further assume that, at equilibrium, any configuration of scars will fully conform to the icosahedral symmetry in all three shapes studied \footnote{Namely, the discrete symmetry group for these crystals is the full achiral icosahedral group $\mathcal{I}_h$, isomorphic to the direct product of the alternating group $A_5$ with point inversions}.

\begin{figure}[t]
\centering  
\includegraphics[width=0.9\columnwidth]{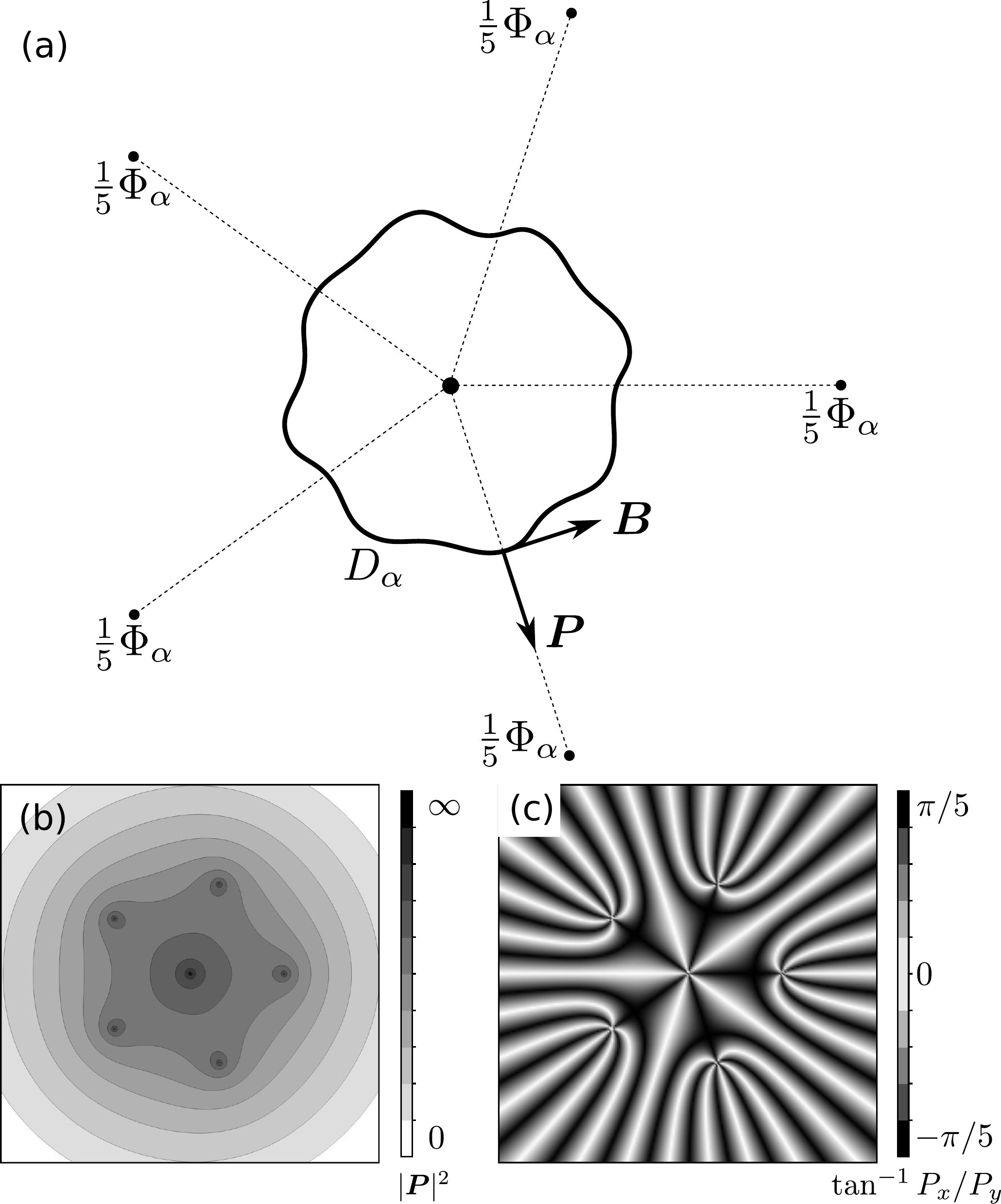}
\caption{\textbf{Effective charge distribution for a cloud of scars around a single seed disclination}. (a) Schematic of the \textit{in-plane} scar-induced charge distribution in our model. The presence of scars can be effectively accounted for by introducing an additional negative charge $-\Phi_\alpha$ at sources, and, due to the icosahedral symmetry of the disclination distribution, five positive charges $\Phi_\alpha/5$ at the sinks. The solutions of \Eqref{eq:rho_curlB} for a flat plane allow us to find analytically the polarization field $\bm{P}$ for the charge distribution in $(a)$. The lower panels show (b) its magnitude and (c) its direction (note that the configuration field is invariant under $2\pi/5$ rotations, so we distinguish angles only $\mathrm{mod}\; 2\pi/5$).}
\label{fig:pentapole}
\end{figure}

While the discrete nature of the disclination charge density \Eqref{eq:rhos_discl} is protected by the topological constraint, dislocations can be treated as a continuum field in the limit $R \gg a$, where several long scar lines are expected. This is particularly advantageous in the case of a large number of excess dislocations, for which calculations based on a discrete description become intractable. 
With this picture in mind, we express the dislocation density via a local Burgers vector density field $\bm{B}=\bm{B}(\bm{r})$, as well as its perpendicular vector field $\bm{P}=\bm{P}(\bm{r})$, representing the dislocation polarization density (i.e. $P^{i}=\epsilon^{ij}B_{j}$, with $\epsilon^{ij}$ the Levi-Civita tensor), thus:
\be
\rhosloc = \nabla \times \bm{B} = - \nabla\cdot\bm{P} \,.
\label{eq:rho_curlB}
\ee
The Burgers vector density must be invariant under the icosahedral symmetry and can therefore be decomposed into the linear superposition of twelve distinct contributions, each corresponding to a specific seed disclination: $\bm{B}=\sum_\alpha \bm{B}_\alpha$. 
\Eqref{eq:rho_curlB} implies that $\rhosloc$ is an exact differential and thus its integral over the full surface vanishes.
On the other hand, integrating over a small crystal patch $D_\alpha$ centered around $\alpha$ and not containing any other disclination, yields the flux of the outgoing scars across the boundary $C_\alpha = \partial D_\alpha$ (see Fig.~{\ref{fig:pentapole}a)}:
\be
\int_{D_\alpha} \dA \nabla \times \bm{B} = - \oint_{C_\alpha} {\rm d}s\,\bm{n}\cdot\bm{P} = - \Phi_{\alpha}\;
\label{eq:circularB}
\ee
where ${\rm d}s$ is the arc-length element along $C_{\alpha}$ and $\bm{n}$ the tangent vector normal to $C_{\alpha}$. Furthermore, since all scars source essentially from the seed disclination $\alpha$, for sufficiently small $C_\alpha$ the flux $\Phi_\alpha$ will be a constant independent on the shape of the contour. Such a field can be generated only by a curl which is singular at the disclination, i.e. $\nabla \times \bm{B}_\alpha= -\Phi_{\alpha}\delta(\bm{r}-\bm{r}_{\alpha})$.
We thus see that, in our description, the main effect of scars is the screening of a seed disclination charge, resulting in a net angular deficit $\pi/3\,q - \Phi_\alpha$ localized at $\bm{r}_\alpha$. 

Now, as the cloud of screening dislocations is topologically neutral over the scale of the whole system (i.e. $\int \dA \rhosloc=0$),  the flux of scars emanating from a given seed disclination must eventually terminate somewhere. In our continuum picture, this implies the existence of field {\em sinks}, whose position is denoted $\bm{r}_{\gamma}$, in proximity of which $\nabla \times \bm{B}_\gamma = \Phi_{\gamma} \delta(\bm{r}-\bm{r}_{\gamma})$.
These sinks can be interpreted as the locus of the terminating end of scars and their effective charge counterbalances the excess disclination screening (see Fig.~\ref{fig:pentapole}a).

If we assume that the icosahedral symmetry of the seed disclinations is inherited by the scar distributions, we have $\Phi_\alpha=\qPhi$ for all sources, each of which is neutralized by the charges at the corresponding scar sinks. For surfaces with positive Gaussian curvature, we expect that $0< \qPhi < (\pi/3)q$, and hence the sinks $\gamma$ would tend to be as far as possible from any given seed disclination. We therefore assume that the scar sinks are located at the centroid of the Delaunay triangles connecting each triplet of disclinations (see \fref{fig:ModelScheme}c). In an perfect icosahedron these locations correspond to the centers of the flat faces. On a sphere, these are equivalent to the vertices of an inscribed regular dodecahedron, the dual solid of an icosahedron. We note that this construction can be thought as a particular case of the so called ``pentagonal buttons'' described in Refs.~\cite{BNT2000,Bowick2006}.

With the model outlined above, we find a local effective screening of disclination stress around the icosahedral vertices, at the price of generating excess defect charges that introduce stress around the face centers. The resulting dislocation density is then parametrized in terms of a single dimensionless quantity, the flux of screening scars $\qPhi$. The topological charge density resulting from the cloud of dislocations screening the twelve disclinations is thus:
\be
\rhosloc = 
\qPhi 
\left[
\frac{3}{5}
\sum_{\gamma=1}^{20} \delta(\bm{r}-\bm{r}_{\gamma})
- 
\sum_{\alpha=1}^{12} \delta(\bm{r}-\bm{r}_{\alpha}) 
\right]
\,,
\label{eq:effective rho}
\ee
where the first sum runs over the sink positions and the second over the seed disclinations sourcing the scars. The additional factor in the first term of \Eqref{eq:effective rho} originates from three sink charges, each having topological charge $\qPhi/5$ (see Fig.~\ref{fig:pentapole}a), collapsing onto a single point.

\subsection{Form of the solutions}
\label{sec:solutions}

We can now rephrase the calculation of the dimensionless stress, \Eqref{eq:laplace_sigma_rhos}, in terms of the effective charge of the seed disclinations and additional charges at the scar sinks,
\be
\nabla^2\sigma
=
\frac{\pi}{3}\sum_{\alpha=1}^{12} \qeff_\alpha \delta(\bm{r}-\bm{r}_\alpha)
+
\frac{\pi}{3}\sum_{\gamma=1}^{20} \qeff_{\gamma} \delta(\bm{r}-\bm{r}_{\gamma})
-
K
\,,
\label{eq:laplace_sigma_eff}
\ee
where
\begin{subequations}\label{eq:qEffs}
\begin{align}
    \qeff_\alpha &= 1-\frac{\qPhi}{\pi/3}\, ,\\
    \qeff_\gamma &= \frac{3}{5}\,\frac{\qPhi}{\pi/3}\,.
\end{align}
\end{subequations}
The equation above implies that $\sigma$ depends linearly on the effective topological charges, given that $\nabla^2$ is a linear operator. Since the stretching energy \Eqref{eq:Estretch} is a quadratic functional of the stress, then $\Estretch$ is necessarily a quadratic polynomial in $\qPhi$. In fact, we can write the energy as 
\be
\Estretch =  \frac{YR^{2}}{2}\left(c_0 + c_1 \qPhi  +  c_2 \qPhi^2 \right) \,,
\label{eq:geometric coefficients}
\ee
where the $c_n$ constants are dimensionless and contain the explicit dependence on the architecture of the lattice (size, shape and effective defect positions). We therefore refer to these as the \textit{geometric coefficients} of the crystal. The problem of finding the stretching energy of a crystal is now translated into calculating the geometric coefficients for specific scar sinks on a given surface. This enormously reduces the level of complexity compared to explicitly considering many discrete dislocations. By expressing the energy in terms of the solutions to \Eqref{eq:laplace_sigma_eff}, we find that the coefficients $c_n$ can be computed from the Laplace-Beltrami spectrum of the manifold (see Appendix \ref{app:spectral} for an explicit derivation), namely: 
\begin{subequations}
\begin{align}
c_0 &= \frac{A}{R^2}
\sum_{n\geq 1}
\frac{1}{\lambda_n^2}
\left|
\frac{\pi}{3A}
\sum_{\alpha=1}^{12} \psi_n(\bm{r}_\alpha)
-
k_n
\right|^2\,, \label{eq:c0}\\
c_1 &= \frac{1}{R^2}\sum_{n\geq 1}
\frac{1}{\lambda_n^2}
\left[
\frac{3}{5} \sum_{\gamma=1}^{20} \psi_n(\bm{r}_{\gamma})
-
\sum_{\alpha=1}^{12} \psi_n(\bm{r}_\alpha) 
\right] \times \nonumber \\
&
\quad\quad
\left[
\frac{\pi}{3A}
\sum_{\alpha=1}^{12} \psi_n^*(\bm{r}_\alpha)
-
k_n^*
\right] \,, \label{eq:c1}\\
c_2 &= \frac{1}{AR^2} \sum_{n\geq 1}
\frac{1}{\lambda_n^2}
\left|
\frac{3}{5} \sum_{\gamma=1}^{20} \psi_n(\bm{r}_{\gamma})
-
\sum_{\alpha=1}^{12} \psi_n(\bm{r}_\alpha) 
\right|^2\,, \label{eq:c2}
\end{align}
\label{eq:express coefficients}
\end{subequations}%
where $\lambda_n$ are the Laplace-Beltrami eigenvalues, $\psi_n$ its eigenfunctions and $k_n$ the projection of the Gaussian curvature onto these. The screening of disclination stress by scars for a particular geometry is now fully reflected in the behaviour of the energy \Eqref{eq:geometric coefficients} with the single free parameter $\qPhi$. Having an analytical solution for the geometric coefficients allows us to study this screening mechanism more in detail. 

The stretching energy has a minimum for a positive $c_2$, which is always the case for a closed crystal as seen from \Eqref{eq:c2}, and it occurs at ${\qPhi_{\text{min}}=-c_1/2c_2}$. Given the sign convention in \Eqref{eq:circularB}, $\qPhi$ is defined such that only positive values will effectively screen the disclinations. This implies that the energy will indeed be lowered in the presence of scars if $c_1<0$. The energy, in fact, attains the minimum value 
\be 
\Estretch^{\text{min}} = \frac{YR^2}{2} \left(c_0 -\frac{c_1^2}{4c_2} \right) \,,
\label{eq:minimal energy}
\ee 
where $c_0$ determines the energy in the absence of excess dislocations. Furthermore, the equilibrium value for the effective charges at sources and sinks are respectively
$\qeff_\alpha=1-(3/2\pi)(|c_1|/c_2)$ and  $\qeff_{\gamma}=(9/10\pi)(|c_1|/c_2)$, and are thus both determined solely by the ratio $c_1/c_2$.

\begin{table}[t]
	\begin{tabular}[c]{lccc}%
	\hline\hline
		 & $c_2$ & $c_1$ & $c_0$ \\
		\hline
	    \multicolumn{1 }{l}{\textbf{Sphere} ($\mathbb{S}_{2}$)} &&&\\
	    $\asph = 0$&&&\\
		Analytical $L=6$ 		& 0.08991 & -0.1211  & 0.04075 \\ 
		Analytical $L\to\infty$ & 0.1031  & -0.1368  & 0.05270 \\ 		
		Numerical fit         				& 0.0925  & -0.1228 & 0.04628 \\	
	    \hline
	    \multicolumn{1}{l}{\textbf{Rounded icosahedron} ($\mathbb{I}_R$)}&&&\\
	    $\asph = 0.0012$&&&\\
		Numerical fit         				& 0.1088 & -0.0386 & 0.00612  \\	
	    \hline
	    \multicolumn{1}{l}{\textbf{Sharp icosahedron} ($\mathbb{I}_S$)}&&&\\
	    $\asph = 0.0020$&&&\\
		Numerical fit        				& 0.1243 & -0.00534 & 0.000481  \\	
		\hline\hline
	\end{tabular}
\caption{Geometric coefficients of the stretching energy \Eqref{eq:geometric coefficients} for the three shapes studied.
The coefficients of the sphere can be calculated analytically with arbitrary precision from \Eqref{eq:sphere energy} by truncating the summation up to mode $\ell=L$. We show both the $L=6$ and $L\to\infty$ values. The numerical results are obtained as described in Appendix~\ref{app:numerical} and correspond to an average of the fitted coefficients for five different crystal sizes.}
\label{table:cCoefficients}
\end{table}

\section{\label{sec:screening}Screening of topological defects}

Having found the expressions for a generic closed crystal, it remains to calculate the geometric coefficients for the three shapes of interest (Fig.~\ref{fig:ModelScheme}). We can then study how a particular scar configuration can effectively screen the disclinations and how this is affected by local curvature. 

For a generic parametric surface, the spectrum of the Laplace-Beltrami operator can be calculated using isothermal (or conformal) coordinates \cite{Bowick2009}. This is particularly straightforward in the case of simple geometries like the sphere, where the calculation is further simplified by the fact that the Gaussian curvature is constant everywhere.  
In this case, an analytical solution of \Eqref{eq:laplace_sigma_eff} can be found using the spherical harmonics, as described in Appendix~\ref{app:spectral_sphere} (see also Ref.~\cite{BNT2000}). 

For the icosahedral shapes, and in general for any shape that cannot be expressed parametrically, we have instead to rely on numerical results. Numerical solutions of \Eqref{eq:laplace_sigma_rhos} are found via a gradient minimization on a triangulation of the surface (for more technical details see Appendix \ref{app:numerical}). We use the analytical results for the sphere as a benchmark for our numerical implementation. The calculated geometric coefficients for these shapes are shown in Table~\ref{table:cCoefficients}.

\begin{figure*}[th]
\centering
\includegraphics[width=0.98\textwidth]{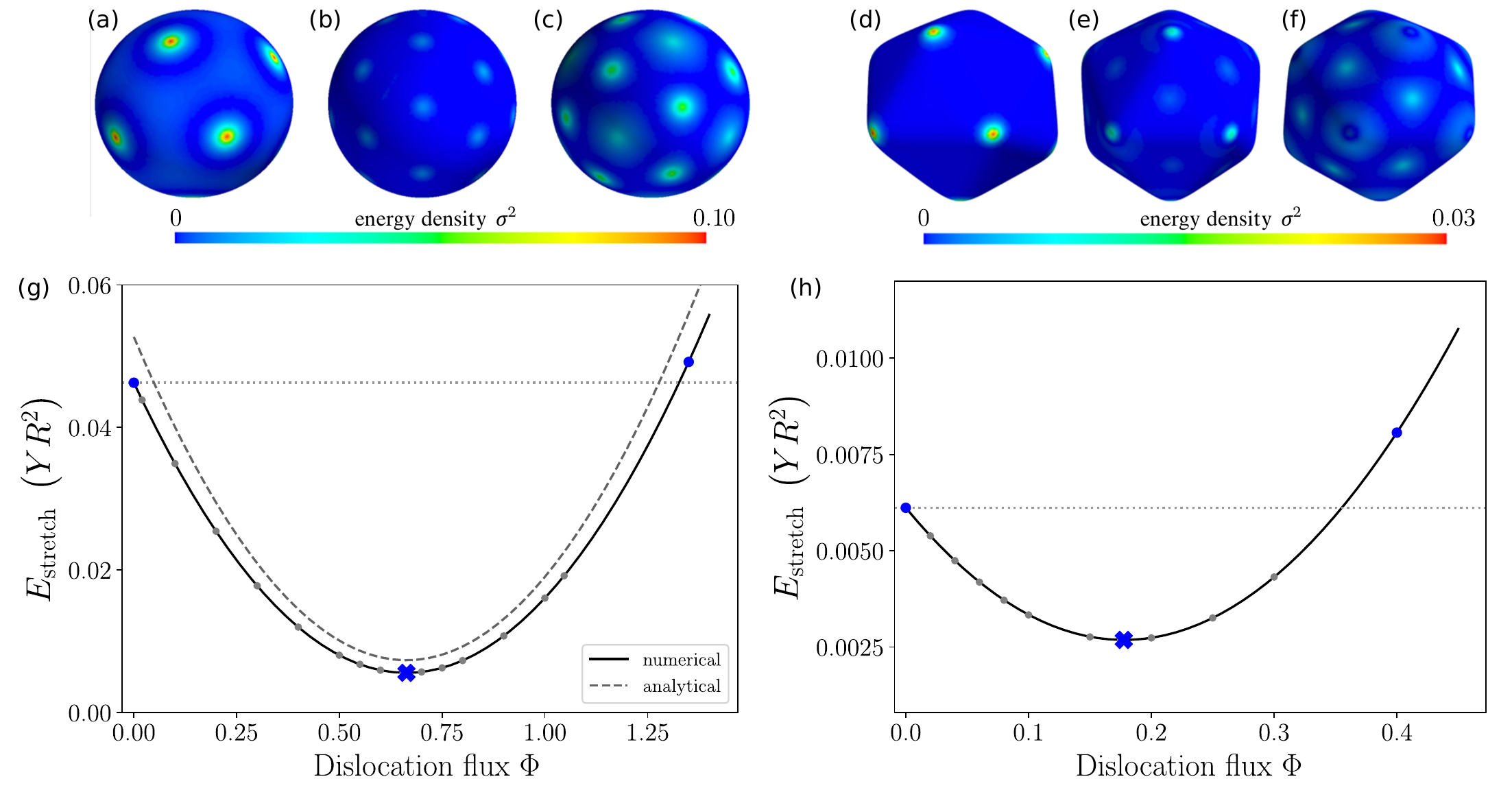}
\caption{\textbf{Screening of disclination stress by dislocation scars in $\mathbb{S}_{2}$ and $\mathbb{I}_R$ of the same size.}
The upper panels show the local energy density on  $\mathbb{S}_{2}$ (a-c) and  $\mathbb{I}_R$ (d-f) for different values of scar density. The lower panels \mbox{(g,h)} show the energy landscape for the two surfaces, with the blue markers corresponding to the snapshots (a-f) and the gray markers the numerical solutions. The solid black line is the best fitting parabola to these values. For the sphere we also show the analytic solution $L\to \infty$ of \Eqref{eq:sphere energy} as a dashed line in panel (g). Note the significant decrease of stress at the seed disclinations for (b), where all effective charges become equal. Conversely, the minimum-energy configuration for $\mathbb{I}_R$ has still a significant difference of charges between sources and sinks.}
\label{fig:Results_numerical_all3}
\end{figure*}	

\subsection{Screening by curvature}

We briefly comment of the scenario where there are no excess dislocations in the crystal, i.e. with  $\qPhi=0$. 
In this case, the stretching energy is simply given by $\Estretch^{\min} = YR^2\,c_0/2$, depending on only one geometric coefficient. The screening of stress by curvature is evident in Table~\ref{table:cCoefficients}, from the decrease of $c_0$ with increasing Gaussian curvature around the seed disclinations, reflected by $\asph$. Indeed, stress is relieved by non-spherical shapes \cite{Funkhouser2013,Lidmar2003,Kohyama2007}. For the sharp $\mathbb{I}_S$, the energy is even reduced by two orders of magnitude compared to the sphere $\mathbb{S}_{2}$, confirming the important role of curvature in reducing disclination-induced strain. 

\subsection{Screening by dislocation scars}

Consider now a crystal with dislocation scars and parametrized by ${\qPhi \ne 0}$. For all three geometries studied, we find that $c_1<1$, showing that disclination stress can be screened out by the  particular scar configuration outlined above. Recall that $\qPhi$ is proportional to the number of scars and hence to the number of excess dislocations. In Fig.~\ref{fig:Results_numerical_all3} we show the effect on the local stress and the energy \Eqref{eq:geometric coefficients}, as more defects are added to the lattice for two particular geometries of the same size: the sphere $\mathbb{S}_{2}$ and the rounded icosahedron $\mathbb{I}_R$. In the upper panels~{(a-f)}, we plot onto the surface an intensity map of the stretching energy density $\sigma^2$ [see \Eqref{eq:Estretch}] for three different values of $\qPhi$: \mbox{(a, d)} no scars, $\qPhi =0$; \mbox{(b, e)} equilibrium, $\qPhi =\qPhi_{\text{min}}$; and \mbox{(c, f)} large $\qPhi$ for which the energy is higher than having no dislocations at all. Note that red denotes regions of high stress, keeping in mind that the color scale is the same for all three plots within each geometry. In panels~\mbox{(g, h)} we plot the stretching energy, \Eqref{eq:geometric coefficients} versus the dislocation flux and highlight with blue markers the three systems in the upper panel. The horizontal dotted line in this plot corresponds to the value of the energy at $\qPhi=0$, namely when there are no scars. In the case of the sphere, the dashed line in Fig.~{\ref{fig:Results_numerical_all3}g} corresponds to the solution for the analytically calculated coefficients \Eqref{eq:express coefficients}. 

\begin{figure}[t]
\centering  \includegraphics[width=\columnwidth]{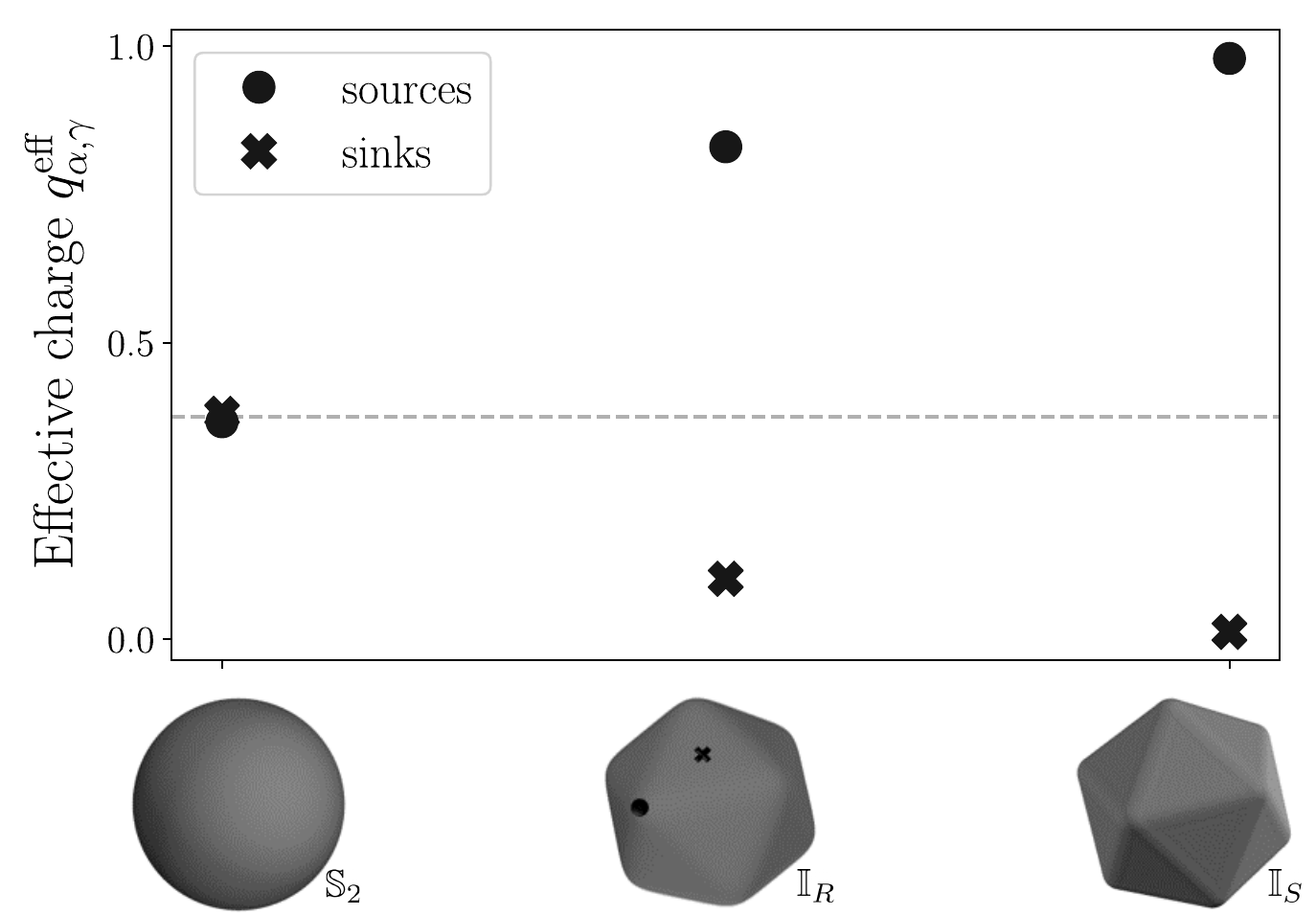}
\caption{\textbf{Effective defect charge at the scar sources $\alpha$ and sinks $\gamma$ for the three shapes considered}. The elastic energy \Eqref{eq:express coefficients} on the sphere is minimized when the 32 defects sitting at the vertices of a rhombic triacontahedron have the same charge $\qeff=3/8$. The flat faces of the icosahedral geometries penalize the sinks, resulting in a higher charge at the vertices.}
\label{fig:Results_effQ}
\end{figure}

Both curves show the same behavior, with a small deviation due to the numerical approximation for the delta functions (details in Appendix~\ref{app:num_integration}). Interestingly, the ratio $c_1/c_2$ is the same for both solutions and thus $\qPhi_{\text{min}}$ is too. This is also shown in Fig.~\ref{fig:Results_effQ}, where we plot the effective charge at the minimum for all shapes. In particular for the sphere, it is worth noticing that the minimal energy is attained for $\qPhi_{\text{min}}=5\pi/24$, i.e. the value for which the effective charge at the disclinations and at the sinks become identical. Since the energy is quadratic in the charges and the crystal has a uniform Gaussian curvature, the stress is equally distributed between sources and sinks. This equilibrium configuration is equivalent to a crystal with 32 point disclinations of charge $\qeff=3/8$, located at the vertices of the inscribed rhombic triacontahedron (which still has the same symmetry group of an icosahedron). The screening is also clearly seen in the surface plot of the energy density in Fig.~\ref{fig:Results_numerical_all3}b.

Instead, icosahedral crystals have very different effective charges at the seed disclinations and at the sinks for the minimum energy configuration, as shown in \fref{fig:Results_effQ}. The higher Gaussian curvature at the vertices can screen more effective charge of the sources as compared to the high cost of the charges of the sinks at the flatter faces (see for example Fig~\ref{fig:Results_numerical_all3}f).  
The effect of higher curvature is also clearly seen in the values of $\qPhi$ at the minimum: for $\mathbb{I}_R$ $\qPhi_{\mathrm{min}} \simeq 0.17$, a value much smaller than $5\pi/24\simeq 0.65$ at the sphere. Even less excess dislocations are expected to form for sharper geometries (for $\mathbb{I}_S$ the equilibrium defect charge is essentially zero when compared to the other geometries), indicating that scars then become less important as a screening mechanism of elastic stress.
Notice further that the magnitude of the energy itself is much smaller in the icosahedron, evident in the scale of the vertical axes of Fig.~\ref{fig:Results_numerical_all3}g and h.

\subsection{The role of crystal size}

Looking back at the scaling behaviour of the equation for the dimensionless stress, \Eqref{eq:laplace_sigma_rhos}, we see that the Laplacian term, $\rhoclin$ and $K$ all scale as an inverse area, namely $\sim 1/R^2$. Conversely, since the Burgers vector $\bm{b}$ is expected to be of the order of the lattice spacing $a$, we must have that the scaling of $\rhosloc$ as $a/R^3$ [see \Eqref{eq:rhos_disloc}]. Consistently, we can factor our this scaling from the dislocation flux by defining $\qPhi\equiv \nD a/R$, where the dimensionless quantity $\nD$ is proportional to the number of excess dislocations $\ndisloc$. In the following, we will refer to it as the {\em dislocation number parameter}.

We can express the energy \Eqref{eq:geometric coefficients} as a function of $\nD$, obtaining the general result
\be
\Estretch = \frac{Y a^2}{2} 
\left[
c_2 \nD^2
+
c_1 \nD\,\frac{R}{a} 
+ 
c_0 \left(\frac{R}{a}\right)^2
\right]\,,
\label{eq:finalenergy}
\ee
which is the main result of this section, as already outlined in the Introduction. This expression holds for any closed crystal of spherical topology and it explicitly shows the size dependence of each term, provided the magnitude of the Burgers vector field is independent of the overall crystal size. 

\begin{figure*}[t]
\centering  \includegraphics[width=\textwidth]{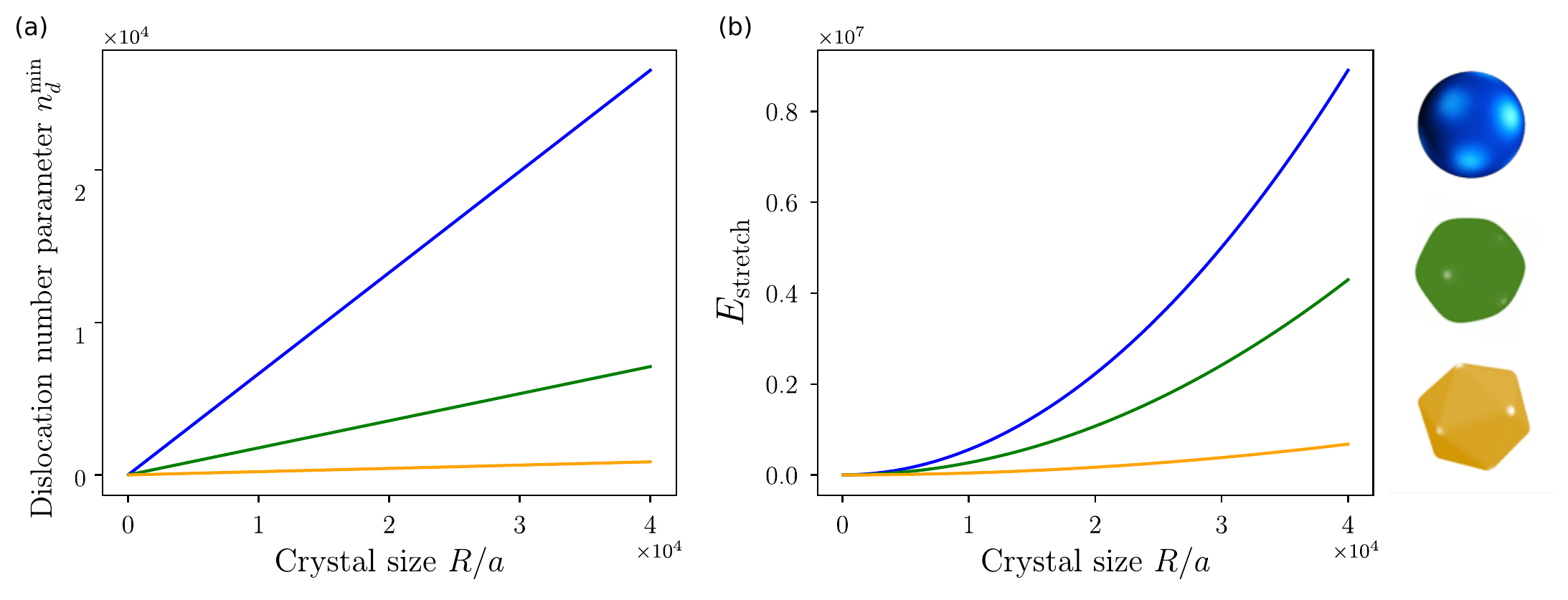}
\caption{\textbf{Equilibrium dislocation number and energy as a function of crystal size.} (a) The number of excess dislocations at the energy minimum grows linearly with the crystal size. This has been previously observed experimentally \cite{Bausch2003}. Dislocations are less relevant for $\mathbb{I}_R$ and even less so for $\mathbb{I}_R$. (b) The quadratic dependence of the stretching energy on the effective charges is translated in a quadratic dependence on crystal size.}
\label{fig:Results_vsSize}
\end{figure*}

Following from the results discussed in the previous section, the dislocation number parameter at the minimum is given by
\be
\nD  = \frac{|c_1|}{2c_2}\frac{R}{a}\,.
\label{eq:nmin_vsR}
\ee 
We thus find that the number of excess defects forming scars grows linearly with the crystal size. This behaviour had been previously reported for spherical crystals in \cite{BNT2000,Bausch2003}. \Eqref{eq:nmin_vsR} shows that this in fact applies to \textit{any} crystal of spherical topology. 
The energy at the minimum then reads
\be 
\Estretch^{\text{min}} = \frac{Ya^2}{2} \left[c_0 -\frac{c_1^2}{4c_2} \right]\left(\frac{R}{a}\right)^2 \,.
\label{eq:minimal_energy_size}
\ee 
Note that the presence of scars does not affect the quadratic scaling of the energy obtained for a crystal having 12 disclinations and no dislocations \cite{BNT2000}. 

The results for the size-dependence of $\nD^{\mathrm{min}}$ and $\Estretch^{\text{min}}$ are shown in Fig.~\ref{fig:Results_vsSize} for the three shapes. For $\mathbb{I}_R$ and $\mathbb{I}_S$ the number of dislocations at equilibrium is much lower than for the sphere. Note however that not even the most curved vertices in $\mathbb{I}_S$ perfectly screen out of the disclination angular deficiency. A finite number of scars are still expected in large icosahedral crystals, even at zero temperature. On the other hand, scars are the most important screening mechanism for the spherical crystal, reflected in the high slope for $\mathbb{S}_{2}$. The differences in the quadratic scaling of the stretching energy \Eqref{eq:minimal_energy_size} are plotted in Fig.~{\ref{fig:Results_vsSize}b}, where we now see clearly the strong effect of curvature as a screening mechanism. In the absence of other energetic terms, closed crystals will always tend towards more icosahedral shapes. In principle, the energy of the spherical crystal could potentially be further lowered by changing the position of the sinks or sources. However, we find this unlikely since we consider a symmetry of scars previously found to be very effective in relieving the stress on spherical crystals. The energy difference between the sphere and the icosahedral shapes implies that a crystal with a low bending modulus would most likely buckle, even in the presence of scars in the lattice. This could explain why buckling has been observed for very large crystals, even when it has typically been identified as a screening mechanism for small ones \cite{Kohyama2007,Guttman2016-2}.

\section{\label{sec:finite_temperature}Dislocation screening at finite temperature}

In this section, we consider the effect of finite temperature $T$ in the organization of the cloud of screening dislocations for the continuum model presented in Sec. \ref{sec:charge_density}. The thermodynamic free energy of a closed crystal is given by $F = \Estretch -TS$, where $S$ is the configurational entropy relative to all possible arrangements of the defects in the crystal. As the number and positions of the seed disclinations is fixed, their associated entropy is constant and equal to $S_{0}=k_{B}\log 12!\approx 20\,k_{B}$. Evidently this contribution has no effect on the configuration of the excess dislocations, thus will be neglected in the following. 
Furthermore, it is safe to assume that only defects which are part of scars will contribute to the free energy. We therefore also neglect the contribution to the entropy from other isolated dislocations that could be thermally induced. In addition, the number of dislocations is very small compared to the total number of lattice points as $N\sim R^{2}$ (see e.g. Fig.~\ref{fig:Results_vsSize}). 

As a first approximation, we calculate the entropy from the number of  possible ways to independently place $\ndisloc$ individual dislocations, ignoring the fact that they form scars. For a small number of defects this yields a higher entropy value, but does not change the general observation that the proliferation of dislocations is entropically favorable. The entropy is then given by $S=k_B\ \ndisloc\ \log \Omega$, where the number of accessible microstates per dislocation is $\Omega=A/A_{d}$, where $A=\int \dA$ is the area of the crystal and $A_{d}\sim a^{2}$ is the area spanned by each dislocation. Now, for the closed surfaces considered here, $A=\zeta R^{2}$, with $\zeta$ a shape-specific coefficient; in particular $\zeta_{\mathbb{S}_{2}}=4\pi$, $\zeta_{\mathbb{I}_{R}}=4.09\pi$ and $\zeta_{\mathbb{I}_{S}}=4.18\pi$. Thus, the shape of the crystals contributes with a $k_{B}\log\zeta$ term to the total entropy.. Being these three contributions small and approximately equal to one another, they will be neglected in the following.
As a result, we can write the free energy as 
\begin{multline}
\label{eq:FreeEcrystal_entropy}
F = \frac{Ya^2}{2}\left[c_2\ \nD^2+c_1\nD\,\frac{R}{a}+c_0\left(\frac{R}{a}\right)^2\right]\\
-\cT T\ \nD\ {\text{log
}}{\left(R/a\right)}\,,
\end{multline}
by taking the stretching energy in \Eqref{eq:finalenergy} and recalling that the number of dislocations is proportional to our model parameter $\nD$. We collect this proportionality and $2k_B$ into a single constant, $\cT$. 

\subsection{Equilibrium crystal configuration}

Because the entropy also depends on the number of dislocations in the crystal, now the equilibrium value of $\nD$ is determined by the minimum of the free energy. As compared to the zero temperature minimum \Eqref{eq:nmin_vsR}, we find an additional sublinear contribution, namely
\begin{equation}
\label{eq:nmin_entropy}
\nD^{\text{min}} =  - \frac{c_1}{2c_2}\ \frac{R}{a}+\frac{1}{c_2} \frac{\cT T}{Ya^2} \ \text{log}\left(R/a\right).
\end{equation}
An addition of a dislocation in the lattice will incur in an elastic cost that opposes the entropic gain. Hence an increase $\nD$ is weighted by the ratio $T/Y$. Also note the geometric coefficient $c_2$ in the second term, hinting at an implicit dependence of the equilibrium configuration on the crystal shape, even when these factors were not explicitly considered in the entropy. For low temperatures, the number of excess dislocation is enough to minimize the stretching energy. We expect the Young modulus to decrease or at least stay roughly constant with increasing temperature, allowing thermally induced dislocations.

\begin{figure*}[t]
\centering  \includegraphics[width=\textwidth]{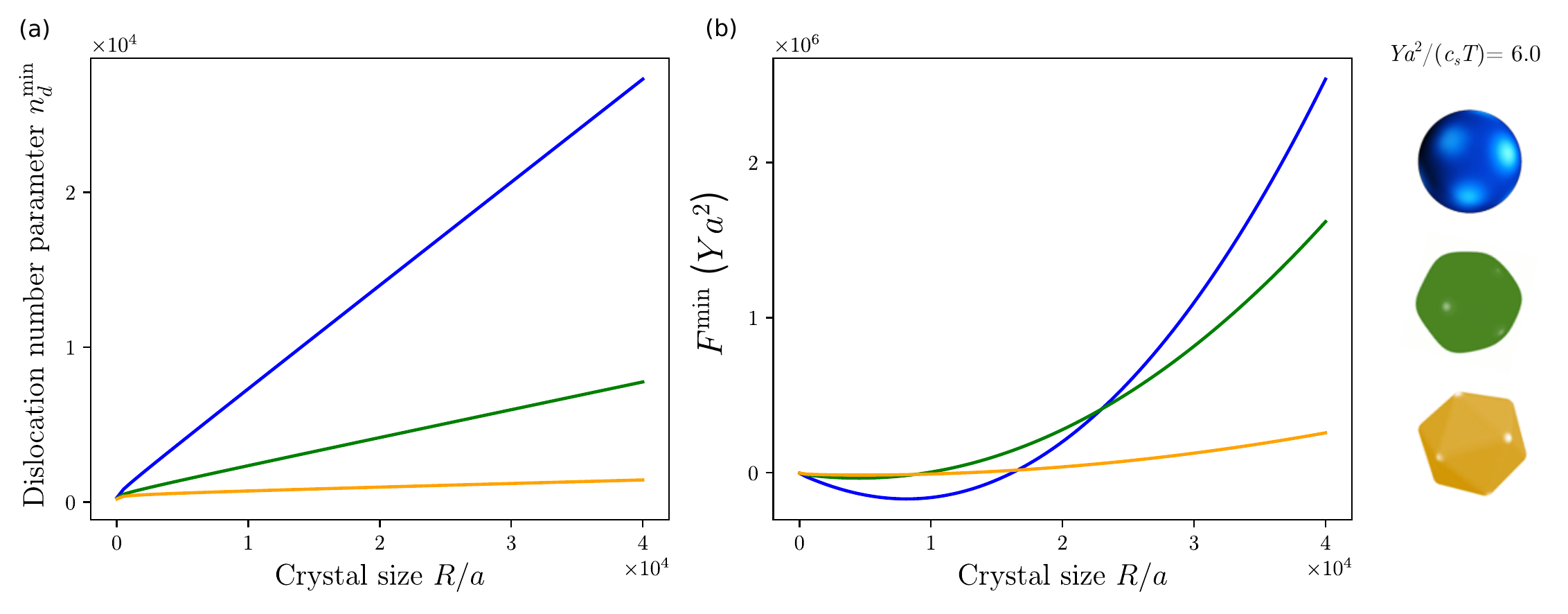}
\caption{\textbf{Equilibrium dislocation number and free energy as a function of crystal size for finite temperature.} For comparison, the expressions for a system at $Ya^2/(\cT T)=6.0$ are plotted for the same range of crystal sizes presented in Fig.~\ref{fig:Results_vsSize}. (a) The behaviour of the number of excess dislocations at the minimum is only changed from the system at zero temperature for very small crystal sizes and barely noticeable. (b) However, there is a remarkable difference in the free energy for small crystal sizes where now dislocations are the main screening mechanism, instead of the increased curvature of the icosahedra.}
\label{fig:Results_vsSize_entropy}
\end{figure*}

Replacing $\nD^{\text{min}}$ back in \Eqref{eq:FreeEcrystal_entropy}, we find the following free energy at the minimum 
\begin{multline}
F^{\text{min}} =  \frac{Ya^2}{2}\left[c_0-\frac{c_1^2}{4c_2} \right]\left(\frac{R}{a}\right)^2\\  
-\cT T\ \text{log}\left(R/a\right)\left[\frac{1}{c_2}\frac{\cT T}{Ya^2}\text{log}\left(R/a\right)-\frac{c_1}{2c_2}\ \frac{R}{a} \right]\,.
\label{eq:Fmin_prefinal}
\end{multline} 
The first term in the sum can be identified as the stretching energy minimum at $T=0$, see \Eqref{eq:minimal_energy_size}. The expression in the squared brackets in the second term has exactly the same form of the equilibrium dislocation number parameter, Eq.~(\ref{eq:nmin_entropy}), but with lower temperature $T\to T/2$. We label this as $\tilde{n}_{d}^{\text{min}}$, which we can use to express \Eqref{eq:Fmin_prefinal} in a more compact form: 
\begin{equation}
\label{eq:Fmin_final}
F^{\text{min}} =  \Estretch^{T=0}- \cT T\ \tilde{n}_{d}^{\text{min}}\ \text{log}\left(R/a\right)\,.
\end{equation}
Once again, we observe the differences between the three shapes with increasing curvature screening around the disclinations. The results are plotted as a function of size in Fig.~\ref{fig:Results_vsSize_entropy}, for the same range of values studied for $T=0$. Looking back at Table~\ref{table:cCoefficients}, we see that the inverse $1/c_2$ decreases for sharper vertices. This means that for the same temperature, the entropy induces a relatively larger number of additional dislocations on a sphere than it does on an icosahedron.
In contrast with the scaling behaviour of the equilibrium dislocation number, we find that the entropy has a strong effect on the equilibrium free energy (see Fig.~\ref{fig:Results_vsSize_entropy}b). At large sizes, stretching plays the main role in the equilibrium configuration and hence curvature is still the most efficient stress screening mechanism. However, for intermediate crystal sizes, icosahedral shapes pay a much higher toll for the additional temperature-induced dislocations. Contrary to Fig.~\ref{fig:Results_vsSize} where high curvature yields always the lowest energy, at finite temperatures dislocation scars are preferred. Hence, for intermediate sizes, spherical crystals with dense dislocation scars are better at screening out the disclination stress compared to buckled icosahedra. 

\section{\label{sec:conclusions}Conclusions}

In this article we introduced a continuum model for dislocation screening in crystalline monolayers with spherical topology. These crystals are naturally found in various experimental set-ups, such as viral capsids~\cite{Caspar1962,Aznar2012}, fullerenes~\cite{Cao2018}, spherical colloidosomes~\cite{Vogel2015,Bausch2003} and surface-frozen emulsion droplets~\cite{Guttman2016-1,Denkov2015}. As a consequence of the spherical topology, these systems necessarily feature a number of disclinations, which in turn, give rise to long-ranged elastic stresses, resulting from the departure of the local coordination number from the ideal $6-$fold configuration of a perfect triangular lattice~\cite{Nelson2002Defects}.
Although never exactly canceled, these stresses can be relieved in two different ways, depending on the density of the underlying lattice, namely the ratio between the typical lattice spacing $a$ and the system size $R$. For spherical crystals with $a\approx R$, 
the Gaussian curvature alone is sufficient to compensate for the angular deficit associated to each disclination and the lowest energy configuration consists of twelve isolated $5-$fold defects placed at the vertices of a regular icosahedron. As the lattice size increases, crystals have been observed to buckle to increase the curvature screening \cite{Lidmar2003}.
Conversely, for $a \ll R$, the crystal is essentially flat at the length scale of the lattice spacing and the underlying Gaussian curvature is no longer sufficient. Note however that buckled geometries have been observed even at this scale \cite{Guttman2016-1, Kohyama2007}. Additional screening can instead be achieved by the proliferation of ``clouds'' of dislocations, whose effect is to delocalize the net topological charge of an isolated disclination on a larger area~\cite{BNT2000,Travesset2003}. Screening dislocations are themselves organized in scars, i.e. chains of alternating $5-$ and $7-$fold disclinations radiating from a given ``seed'' disclination \cite{Bausch2003,Guerra2018}.

Unlike grain boundaries in planar crystals that propagate across an infinite length in open systems, scars in boundaryless, closed crystals starting in the proximity of seed disclinations terminate within the lattice. This peculiarity gives rise to fluxes of outgoing and incoming $5-7$ dislocation dipoles in specific regions of the crystal, depending on the location of the seed disclinations. The approach proposed here takes advantage of this property to capture the effect of dislocation screening of the stress in a coarse-grained fashion. Upon treating the dislocation flux itself as an independent degree of freedom, we have demonstrated that the calculation of the stretching energy of the crystal can be significantly simplified and reduced to the evaluation of just three {\em geometric coefficients}, solely related with the geometry of the underlying surface via the spectrum of the Laplace-Beltrami operator and the projection of the Gaussian curvature.

Consistent with experimental observations \cite{Bausch2003,Einert2005}, we found that the number of excess dislocations increases linearly with the system size, and that this is a general feature of any crystal with spherical topology. While the latter result can also be obtained from simple scaling arguments, our approach allows to estimate the prefactors of such a scaling relations. We can therefore investigate the effect of curvature, focusing on the performance of dislocation screening and can demonstrate that, for any finite curvature, dense icosahedral crystal will always feature a cloud of screening dislocation in proximity of the topologically required seed disclinations. 
Additionally, we study the effect of entropy on the equilibrium configuration of crystals at finite temperature. We show that this equilibrium can be expressed as the zero temperature minimum plus a single temperature-dependent term. We find that although entropy tends to always favor denser scars, temperature induces new dislocations on a crystal in a geometry-dependent fashion. 

Our approach offers evident advantages compared to the traditional discrete treatment of individual dislocation dipoles \cite{BNT2000}. For dense systems, where the number of screening dislocations is large, a discrete treatment becomes computationally prohibitive, whereas our method highly reduces the degrees of freedom from all individual defect positions to only those of the seed disclinations and the terminating end of the scars.
The calculation of the lowest energy configuration and their associated stretching energy is simplified to the computation of three geometric coefficients, for which we outline a numerical implementation that can be used, in principle, for arbitrary shapes. 
These coefficients do not depend on the density of dislocation scars nor on the crystal size. Therefore, our method allows to readily analyze systems of different sizes and with varying number of excess dislocations, starting from a single well-invested calculation. Furthermore, other contributions to the total elastic energy, such as bending, can be straightforwardly incorporated into the model without compromising the simplicity of this approach.

Finally, to assess the efficiency of our method, we have considered three examples of crystals of spherical topology: a sphere and two rounded icosahedra with different sharpness at corners and edges. When the bending cost in the crystal is negligible we find that curvature is the main screening mechanism of disclination stress, at zero temperature, therefore resulting in buckled geometries. However, at finite temperatures entropy leads to denser dislocations scars in spherical crystals compared to the icosahedral shapes, which in turn can stabilize the spherical shape at intermediate crystal sizes. For sake of the presentation, we considered the simplest case of crystals with icosahedral symmetry, for which the flux of screening disclinations is uniform across the surface and could ultimately be modeled in terms of a single dimensionless number, $\qPhi$. For less regular geometries, $\qPhi$ is a generic scalar field, whose equilibrium configuration could be found upon minimizing the total elastic energy. 

\acknowledgements
	
We are grateful to Eli Sloutskin for several interesting conversation that inspired this research. This work is partially supported by Netherlands Organisation for Scientific Research (NWO/OCW), as part of the D-ITP program (I.G.A.), the Vidi scheme (P.F. and L.G.) and the Frontiers of Nanoscience program (L.G.).

\appendix
\renewcommand{\thesubsection}{\Alph{section}.\arabic{subsection}}

\section{Closed crystals with arbitrary sinks and sources}
\label{app:analytical}

\subsection{Solutions}

\label{app:spectral}

The resulting equation for the dimensionless stress $\sigma$ for our model, \Eqref{eq:laplace_sigma_eff} can be written in the more general form
\be
\nabla^2 \sigma(\bm{r}) = \frac{\pi}{3}\sum_{j} q^{\text{eff}}_j \delta(\bm{r}-\bm{r}_j) - K(\bm{r}) \,,
\label{eq:general structure}
\ee
for an arbitrary distribution of disclinations of topological charge $\qeff_j$. Every scalar function on a compact manifold can be expressed as a unique linear combination of the countable eigenfunctions of the Laplace-Beltrami operator, generalizing the notion of Fourier and spherical harmonics to arbitrary (closed) manifolds (i.e. Sturm-Liouville decomposition). Thus one can write
\be 
\sigma(\bm{r}) = \sum_{n \geq 0} \sigma_n \psi_n(\bm{r}) \,,
\ee
where $\sigma_n$ are the Sturm-Liouville coefficients for the stress and $\psi_n$ are the eigenfunctions, for which
\be 
\label{eq:OpEigenvalues}
\nabla^2\psi_n(\bm{r}) = \lambda_n\psi_n(\bm{r})\,.
\ee
In general, the spectrum of eigenvalues is degenerate (as for the sphere). Nonetheless, the eigenfunctions $\psi_{n}$ are mutually orthogonal and can be normalized in such a way that
\be
\frac{1}{A} \int \dA \psi_n \psi_m^* = \delta_{nm} \,,
\label{eq:normalization}
\ee
with $A= \int \dA $ the total area of the crystal. This expression implies that the eigenfunctions are dimensionless quantities.
We also write the decomposition of the Gaussian curvature 
\be 
K(\bm{r}) = \sum_{n \geq 0} k_n \psi_n(\bm{r}) \,.
\ee
Note that the eigenvalues are all non-negative, with a single $\lambda_0=0$ and $\psi_0$ a constant function. The zero-th mode of (the trace of) the stress is associated to the amount of stretching due to pre-existing incompatibility in the crystal (i.e. in absence of defects). We assume that, at equilibrium, this \textit{pre-stress} is vanishing, i.e. 
\be 
\sigma_0 = 0 \,.
\ee
The existence of the constant mode and the mutual orthogonality of all other eigenfunctions implies that all $\psi_{n\geq 1}$ have vanishing expectation value over the manifold. It is then easy to solve for every $n$, finding
\be
\sigma_n
= 
\frac{1}{\lambda_n}
\left[ 
\frac{\pi}{3 A}
\sum_j \qeff_j \psi_n(\bm{r}_j)
-
k_n
\right] \,,
\ee
for $n\geq 1$. The equation for $n=0$ (for which $\psi_0=1$ in our normalization) simply restates the topological identity
\be
A \, k_0 = \frac{\pi}{3}\sum_j \qeff_j \,,
\ee 
which also follows from $k_0= 6 \chi/A$, with $\chi$ the Euler characteristic. The total energy at equilibrium can be written as
\be 
\Estretch
=
\frac{Y A}{2}
\sum_{n\geq 1}
\frac{1}{\lambda_n^2}
\left|
\frac{\pi}{3 A}
\sum_j \qeff_j \psi_n(\bm{r}_j)
-
k_n
\right|^2\,,
\label{eq:Estretch decomposed}
\ee
which proves that $\Estretch$ is quadratic in the topological charges. 

For the effective charges considered in \Eqref{eq:qEffs}, the geometric coefficients can be evaluated explicitly. By expanding the solution above and rewriting it in powers of the effective charge $\qPhi$, we find the expressions in \Eqref{eq:express coefficients} in the main text. 

\subsection{Spherical harmonics}
\label{app:spectral_sphere}

For a perfectly spherical crystal with radius $R$, we have that $K=1/R^2$ and thus $k_{n\geq 1}=0$. Furthermore, the degeneracies of the spectrum are well understood. The eigenfunctions are the spherical harmonics $Y_\ell^m(\theta,\phi)$ (with $\theta$ and $\phi$ the usual azimuth and longitudinal angles), with eigenvalues $\ell(\ell+1)/R^2$, with $\ell\in \mathbb{N}$. The degeneracy of each eigenvalue is then $(2\ell+1)-$fold. Explicitly, the spherical harmonics can be written in terms of Legendre polynomials as 
\be 
Y_\ell^m(\theta,\phi)=\sqrt{\left(2\ell+1\right)\frac{(\ell-m)!}{(\ell+m)!}}\,P_\ell^m(\cos \theta)e^{im\phi} \,,
\ee 
where the normalization (the so-called \textit{geodesic} convention) is chosen such that \Eqref{eq:normalization} is satisfied. With this basis, we have $\psi_\ell(\theta,\phi) = \sum_{m=-\ell}^{\ell} Y_\ell^m(\theta,\phi)$. It is then straightforward to prove that the stretching energy \Eqref{eq:Estretch decomposed} takes the form 
\begin{multline}
\Estretch
=
\frac{\pi Y R^2}{72}
\sum_{\ell=1}^{\infty} \frac{2\ell+1}{\ell^2(\ell+1)^2}\times\\
\sum_{m=-\ell}^{\ell}\frac{(\ell-m)!}{(\ell+m)!}\left|\sum_{j} \qeff_j P_\ell^m(\cos \theta_j) e^{im\phi_j}\right|^2 \,,
\label{eq:sphere energy}
\end{multline}
which, when evaluated on defect distributions invariant under the icosahedral group $\mathcal{I}_h$. This solution matches with previous results reported for a crystal with only twelve $5-$fold disclinations \cite{BNT2000}, for which in our expression it would be equivalent to $\qeff_j=1$. Although this sum is infinite, each addend can be calculated analytically, so that estimates with arbitrary precision are possible. Interestingly, not every harmonic degree contributes to $\Estretch$; it was already argued in Ref. \cite{BNT2000} that only irreducible representations of $SO(3)$ that contain the trivial representation of $\mathcal{I}_h$ give a non-zero contribution to the energy. Furthermore, in Ref.~\cite{BNT2000}, it was found that all odd $l$ never contain the trivial irreducible representation, and the first four non-trivial modes are $\ell=6,10,12,16$. 

For our model with dislocations, the sum in $j$ splits into 12 $\alpha-$terms at the icosahedral symmetry and 20 $\gamma-$terms at the dodecahedral symmetry [see \Eqref{eq:qEffs}]. We calculate the analytical geometric coefficients for the sphere shown in Table~\ref{table:cCoefficients} with the expansion above up to some cut off $\ell=L$. The value for $L \to\infty$ corresponds to $L=500$.

\subsection{More general crystal architectures}

Since we are interested in investigating curvature and scars as screening mechanisms, we focus on crystals with icosahedral shapes and also icosahedral symmetry in the lattice. However, we can take a step back and write our model in a slightly more general fashion. More specifically we can challenge the following assumptions used in constructing our model, and show that our results for the stretching energy still holds: \textit{i)} There are only 12 disclination cores, \textit{ii)} these are regularly spaced with icosahedral symmetry, \textit{iii)} scars for different disclination cores share the same sink position, and \textit{iv)} these sinks have dodecahedral symmetry. We keep the assumption that scars are sourced at $\ndclin$ disclination positions, terminating somewhere far away within the crystal and thus having an associated $\Phi_{\alpha}$ dislocation flux per disclination core. Bound to the topological constraint of zero total dislocation charge, we include $N_{\rm sink}$ sinks for scars, located at some positions $\bm{r}_{\gamma}$. Thus, we write a similar equation for the dimensionless stress, 
\begin{equation}
\nabla^2 \sigma(\bm{r}) = \frac{\pi}{3} \sum_{\alpha}^{\ndclin} \qeff_\alpha \delta(\bm{r}-\bm{r}_\alpha) +  \frac{\pi}{3} \sum_{\gamma}^{N_{\rm sink}}\qeff_{\gamma} \delta(\bm{r}-\bm{r}_{\gamma}) - K(\bm{r}) \,,
\end{equation}
with charge densities given by
\begin{subequations}
    \begin{align}
    \qeff_\alpha &=  q_\alpha-\frac{\Phi_{\alpha}}{\pi/3}\,,\\
    \qeff_{\gamma} &= \frac{\ndclin}{N_{\rm sink}}\,\frac{\Phi_{\gamma}}{\pi/3}\;.
    \end{align}
\end{subequations}
We thus see that the general solution for the stress and the energy holds for this generic lattice architecture too. The number and positions of disclination sources and of scar sinks changes the values of the geometric coefficients $c_n$, hence changing the stretching energy of the crystal. Nonetheless, the considerations on size scaling of the energy and of the number of excess dislocations remain valid also in this case. 

\section{Numerical methods}
\label{app:numerical}

Lacking an explicit parametrization for $\mathbb{I}_R$ and $\mathbb{I}_S$, the Laplace-Beltrami spectrum for these shapes is impossible to obtain analytically. We have to solve numerically the equation for the stress \Eqref{eq:laplace_sigma_eff} via a gradient minimization. We calculate the geometric coefficients by fitting the quadratic energy as a function of different values of the effective charge $\qPhi$. This method can be used to study the mechanics crystals with other closed convex shapes that can be similarly discretized.

\subsection{Meshed geometries for numerical integration}
\label{app:geometries}

The discretization of the sphere and the icosahedra was done using the software Surface Evolver \cite{SurfaceEvolver}. We built a perfectly sharp icosahedron with the 12 vertices connected by straight lines and hence perfectly flat faces, which was refined to have $\sim 2\times 10^4$ mesh points. Using the software, we allow an area-minimizing relaxation of the icosahedron into a sphere, subjected to volume conservation. The icosahedral shapes used in this work, $\mathbb{I}_S$ and $\mathbb{I}_R$ are intermediate shapes in this evolution, while the sphere $\mathbb{S}_{2}$ is the final shape obtained. The size $R$ of crystal is given by the radius of this last geometry.

\begin{figure}[t]
\centering
\includegraphics[width=0.9\columnwidth]{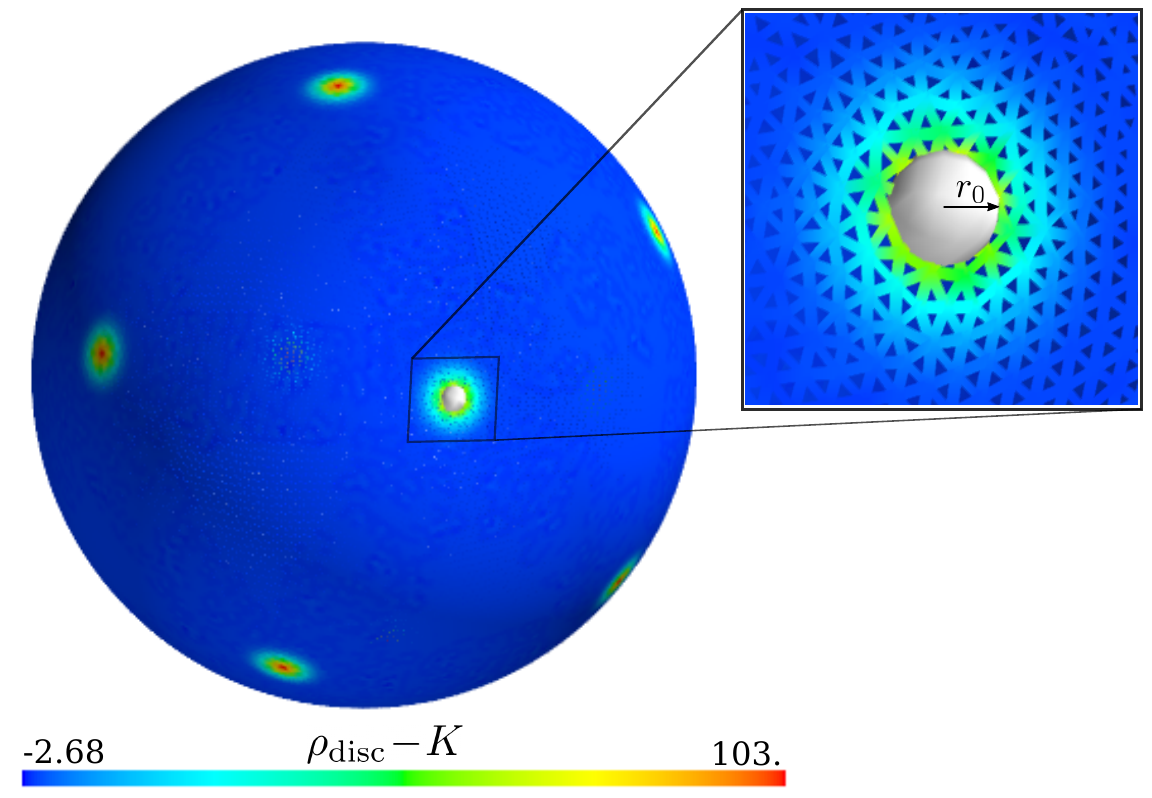}
\caption{\textbf{Triangulation of the sphere and disclination core size.} The mesh consists of 22879 vertices (this number is not related to the number of lattice points in the physical crystal). The tiny white sphere shows the disclination core size used in the numerical solutions. We approximate the Dirac deltas as Gaussian functions of variance $r_0^2$. The color code shows the the local source terms for the stress on the sphere, where $\Phi=0$ and $r_0/R=0.04$.}
\label{fig:app_meshedGeo}
\end{figure}

Note that the triangulation obtained is generally irregular. However the mesh was adjusted in such a way that all triangles have roughly equally long edges
with length dispersion $(l^{\max}-l^{\min})/ \langle l \rangle = 0.48,\ 0.54$ and $0.55$ for $\mathbb{S}_2$, $\mathbb{I}_R$ and $\mathbb{I}_S$ respectively. Since the area variation is not too large, all meshes obtained still have on the order of $10^4$ points. It is important to remark that this discretization is absolutely not linked to the underlying crystal structure. In fact, the triangulation is coarse-grained with respect to the actual lattice so that each triangle encloses a large number of lattice sites.
See Fig.~\ref{fig:app_meshedGeo} for an explicit representation of the triangulation used for the sphere.  

The three shapes are characterized by their asphericity, which can be defined as
\begin{equation}
\asph= \frac{\langle\Delta R^2\rangle}{\langle R^2\rangle} = \frac{1}{N_{\text{vert}}} \sum_{v=1}^{N_{\text{vert}}} \frac{\left(R_v-\langle R \rangle \right)^2}{\langle R^2 \rangle}\,,
\label{eq:asphericity}
\end{equation}
with $R_v$ being the radial distance between each mesh point (or vertex) $v$ to the geometric center, and $\langle R \rangle = 1/N\sum_v^N R_v$. For the particular shapes obtained with Surface Evolver we calculated: for $\mathbb{S}_{2}$, ${\asph = 9\times 10^{-9}}$; for the round $\mathbb{I}_R$, ${\asph = 0.0012}$; and for the sharp $\mathbb{I}_S$, ${\asph = 0.0020}$.
For reference, a perfectly round sphere has zero asphericity and the initial perfectly sharp icosahedron (all flat faces and diverging mean curvature at vertices and edges) has ${\asph = 0.0026}$. 

\subsection{Numerical integration}
\label{app:num_integration}

We solve a discretized version of \Eqref{eq:laplace_sigma_eff} on each mesh point $v$ of surface, given in general by 
\be
\nabla_v^2\sigma_v = \sum_{i=1} \qeff_j \Delta(r_{j,v}) - K_v\,,
\label{eq:laplace_sigma_discrete}
\ee
where $\Delta(r_{i,v})$ is an approximation to the Dirac delta-functions of the original equation that depends only on the geodesic distance between the vertex $v$ and the $i-$th effective disclination, given by ${r_{i,v}=|\bm{r}_j-\bm{r}_v|}$. We take $\Delta(r_{i,v})$ to be a Gaussian function,
\begin{equation}
\Delta(r_{i,v}) = f(r_0)\ \exp \left(-r^2_{i,v}/r_0^2\right)\,,
\label{eq:delta_approx}
\end{equation}
where we refer to the parameter $r_0$ as the \textit{core size} and we take $r_0\gtrsim l^{\max}$ (see Fig.~\ref{fig:app_meshedGeo}). For all three meshes, we took $r_0/R = 0.04$. The Gaussian is normalized by a factor $f(r_0)$ so that the topological constraint in the total defect charge is met, i.e. ${\sum_v\sum_j \qeff_j \Delta(r_{\alpha,v})A_v = 4\pi}$.  The area $A_v$ of each point is given by the Voronoi tessellation of the mesh. The numerical solver based on gradient minimization was built in-house, using C++. The calculation of the geodesic distances on a mesh was done by an implementations of the algorithm \cite{Mitchell1987} available at \cite{Geodesic}. 

We then simply integrate the stretching energy as ${\Estretch=\sum_v \sigma_v^2 A_v}$. For a given shape of size $R$, we calculate the energy for several different values of $q_D$ while keeping the position of the defect cores and $r_0$ fixed (see for example the light gray markers in Fig.~\ref{fig:Results_numerical_all3}). The discretized energy $E(\nD)$ is then fitted with a polynomial of order two, consistent with \Eqref{eq:finalenergy}. Note however that the parameters obtained from the fitting, which we denote $c_n'$, will vary for different $R/a$. The size-dependence is then divided out in order to get the geometric coefficients for that particular geometry. The $c_n$ coefficient for a given shape (shown in Table~\ref{table:cCoefficients}) is the average of the fitted values over different sizes $R$, thus $c_n^{\rm{shape}} = \langle c_n' /R^{2-n}\rangle_R$.

%\bibliography{bitDroplets}{}

%merlin.mbs apsrev4-1.bst 2010-07-25 4.21a (PWD, AO, DPC) hacked
%Control: key (0)
%Control: author (8) initials jnrlst
%Control: editor formatted (1) identically to author
%Control: production of article title (-1) disabled
%Control: page (0) single
%Control: year (1) truncated
%Control: production of eprint (0) enabled
%

\end{document}